# ELEMENTARY PARTICLE PHYSICS AND FIELD THEORY

## THE EFFECTIVE ACTION FOR SUPERFIELD LAGRANGIAN QUANTIZATION IN REDUCIBLE HYPERGAUGES

A. A. Reshetnyak                                                  UDC 539.01

*The rules of local superfield Lagrangian quantization in reducible non-Abelian hypergauge functions are formulated for an arbitrary gauge theory. The generating functionals of standard and vertex Green's functions which depend on the Grassmann variable η via super(anti)fields and sources are constructed. The difference between the local quantum and the gauge fixing action determines an almost Hamiltonian system such that translations with respect to η along the solutions of this system define the superfield BRST transformations. The Ward identities are derived and the gauge independence of the S-matrix is proved.*

## INTRODUCTION

The BRST symmetry principle underlies both the canonical [1] and the covariant quantization scheme [2] for general gauge theories and their superfield generalizations [3–5]. The superfield quantization [3], which is applicable in the canonical formalism and in its implication – Lagrangian formalism, makes use of the nontrivial relation of the odd Grassmann η and even $t$ projections of supertime, $\Gamma = (t, \eta)$, as distinct from the Lagrangian quantization [4, 5].

Note that algorithmic methods have been found [6] for constructing generalized Poisson sigma models in the framework of the superfield formalism [7]. The quantization described in [3] has been generalized for two and more supersymmetries which are associated with Grassmann variables $\eta^1,…,\eta^N$ [8].

The aim of the work under consideration is to construct a local version of the superfield Lagrangian quantization (SLQ). In the SLQ, we realize the explicit superfield representation of the structural functions of a gauge algebra (GA), not indicated in [4, 5], in the framework of an η-local superfield model (SM) which includes the initial standard gauge model.

A correlation with classical mechanics reconstructs the dynamics and gauge invariance for the original model in terms of η-local differential equations (DE's). The properties of the local generating functionals of Green's functions (GFGF's) are derived from a Hamiltonian system (HS), which is constructed w.r.t. the η-local quantum and the gauge fixing action.

In the SLQ, we first define the effective action for a wider class of non-Abelian reducible hypergauges (the case of irreducible hypergauge functions is considered in [9]).

In this paper, we describe the *Lagrangian* and *Hamiltonian formulations* of the SM, specify quantization rules, and determine, based on the component formulation, the relations of the proposed quantization scheme to the superfield quantization [4, 5] and multilevel formalism [9].

We make use of some of conventions from [4, 5] and the condensed notation from [10]. The rank of an even supermatrix is characterized by a pair of numbers $(k_+, k_-)$, where $k_+$ and $k_-$ are the respective ranks of the Bose–Bose and Fermi–Fermi blocks of the supermatrix with respect to the basic Grassmann parity ε. A similar pair of numbers denotes the dimension of a supermanifold, which is equal to $(\dim_+, \dim_-)$. On the set of these pairs, operations of component-wise composition and comparison $((k_+, k_-)>(l_+, l_-) \Leftrightarrow ((k_+>k_-, l_+\geq l_-)$ or $(k_+\geq k_-, l_+> l_-))$, $(k_+, k_-)=(l_+, l_-) \Leftrightarrow ( k_+=l_+, k_-=l_-))$ are defined.

---

          

# THE LAGRANGIAN AND HAMILTONIAN FORMULATIONS OF A SUPERFIELD MODEL

The basic objects in the Lagrangian and Hamiltonian formulations of an SM are the Grassmann-valued $C^\infty(\Pi TM_{cl})$ and $C^\infty(\Pi T^*M_{cl})$ functions of the respective *Lagrangian* and *Hamiltonian actions*[1]:

$$S_L : \Pi TM_{cl} \times \{\eta\} \to \Lambda_1(\eta, R), \qquad S_H : \Pi T^*M_{cl} \times \{\eta\} \to \Lambda_1(\eta, R), \quad \varepsilon(S_L) = \varepsilon(S_H) = \mathbf{0} \tag{1}$$

and the functionals $Z[A]$ and $Z_H[\Gamma_l]$, nonequivalent to these functions, whose densities are defined accurate to the corresponding functions $f(A(\eta), \partial_\eta A(\eta), \eta)$ and $f(\Gamma_l(\eta), \eta) \in Ker\{\partial_\eta\}$, $(\int d\eta = \partial_\eta, \ \partial_\eta \equiv \frac{d_l}{d\eta})$:

$$(Z[A], \ Z_H[\Gamma_l]) = \left(\partial_\eta S_L(\eta), \ \partial_\eta\left(\frac{1}{2}\Gamma_l^p \omega_{pq}^l \partial_\eta^r \Gamma_l^q - S_H\right)(\eta)\right), \quad \varepsilon(Z) = \varepsilon(Z_H) = (1, 0, 1). \tag{2}$$

The values of the Grassmann parities $\varepsilon = (\varepsilon_P, \varepsilon_{\bar{J}}, \varepsilon)$, ($\varepsilon = \varepsilon_P + \varepsilon_{\bar{J}}$), where $\varepsilon_{\bar{J}}$ and $\varepsilon_P$ are auxiliary $Z_2$-gradings w.r.t. the coordinates $z^M$ and $\eta$ of the superspace[2] $M = \tilde{M} \times \tilde{P}$, are defined for $A^i(\eta)$ by the rule $\varepsilon(A^i) = ((\varepsilon_P)_i, (\varepsilon_{\bar{J}})_i, \varepsilon_i)$. The supermatrix $\|\omega_{pq}^l(\eta)\|$ is inverse of the supermatrix $\|\omega_l^{pq}(\eta)\| = \left\|\left(\Gamma_l^p(\eta), \Gamma_l^q(\eta)\right)_\eta\right\|$: $\omega_l^{pq}(\eta)\omega_{qd}^l(\eta) = \delta_d^p$ which is defined in terms of the local superantibracket $(\bullet, \bullet)_\eta = \frac{\partial_r \bullet}{\partial \Gamma_l^p(\eta)} \omega_l^{pq}(\eta) \frac{\partial_l \bullet}{\partial \Gamma_l^q(\eta)}$, where $\frac{\partial_{r(l)}}{\partial \Gamma_l^p(\eta)}$ is the right (left) superfield variational derivative w.r.t. the superfield $\Gamma^p_l(\eta)$ for a fixed $\eta$.

Assuming the existence of a critical superfield configuration for the functionals $Z[A]$ and $Z_H[\Gamma_l]$, we may code the SM dynamics, respectively, by superfield Euler–Lagrange equations, which, in view of the identities $\partial_\eta^2 A^i(\eta) = 0$, are equivalent to a *Lagrangian system* (LS), and by a *Hamiltonian system* (HS)

$$\begin{cases} \partial_\eta^2 A^i(\eta) \dfrac{\partial_l^2 S_L(\eta)}{\partial(\partial_\eta A^i(\eta)) \partial(\partial_\eta A^j(\eta))} = \partial_\eta^2 A^i(\eta)(S''_{L\ ij})(\eta) = 0, \\[2mm] \Theta_l(\eta) = \dfrac{\partial_l S_L(\eta)}{\partial A^i(\eta)} - (-1)^{\varepsilon_i}\left(\dfrac{\partial_l}{\partial \eta}\dfrac{\partial_l S_L(\eta)}{\partial(\partial_\eta A^i(\eta))} + (\partial_\eta U_+)(\eta)\dfrac{\partial_l S_L(\eta)}{\partial(\partial_\eta A^i(\eta))}\right) = 0; \end{cases} \tag{3}$$

$$\partial_\eta^r \Gamma_l^p(\eta) = \left(\Gamma_l^p(\eta), \ S_H(\eta)\right)_\eta, \tag{4}$$

---

[1] Here, $\Pi TM_l = \{(A^i, \partial_\eta A^i)(\eta) \mid A^i(\eta) = (A^i + \lambda^i \eta) \in M_{cl}, \ i = 1, ..., n = n_+ + n_-\}$ and $\Pi T^*M_l = \{\Gamma_l^p(\eta) = (A^i, A_i^*)(\eta) \mid A_i^*(\eta) = A_i^* - \eta J_i, \ l = cl\}$ are the odd tangent and cotangent bundles over the configuration space of classical superfields $A^i(\eta)$ with $n_+$ bosonic and $n_-$ fermionic degrees of freedom with respect to the $\varepsilon$-parity for fixed continuous components of the condensed index $i$, and the quantities $A^i$, $\lambda^i$, $A_i^*$, and $J_i$ for $\varepsilon_P(A^i(\eta)) = 0$ are, respectively, the classical fields, Lagrangian multipliers, antifields, and sources to the fields $A^i$ of the Batalin–Vilkovisky (BV) quantization method.

[2] The superfields $\Gamma^p_l(\eta)$ are defined on $M$, which can be realized as the quotient of a global symmetry supergroup $J$ for the functionals $Z[A]$, $Z_H[\Gamma_l]$, $J = \bar{J} \times P$, $P = \exp\{\imath\mu p_\eta\}$, where $\mu$ and $p_\eta$ are the nilpotent parameter and the generator of $\eta$-shifts, respectively ($\mu^2 = p_\eta^2 = 0$), with a spacetime supersymmetry group chosen for $\bar{J}$.

where the nilpotent operator $(\partial_\eta U_+)(\eta) = \partial_\eta A^i(\eta) \frac{\partial_l}{\partial A^i(\eta)} = [\partial_\eta, U_+(\eta)]_-$, $U_+(\eta) = \eta \partial_\eta A^i(\eta) \frac{\partial_l}{\partial A^i(\eta)}$ is introduced. The equivalence of both formulations and, hence, the coincidence of $Z[A]$ with $Z_H[\Gamma_l]$ and LS (3) with HS (4) is guaranteed by the nondegeneracy of $\|(S''_{L\,ij})(\eta)\|$ in the framework of a Legendre transform of $S_L(\eta)$ with respect to $\partial_\eta^r A^i(\eta)$:

$$S_H(\Gamma(\eta), \eta) = A_i^*(\eta) \partial_\eta^r A^i(\eta) - S_L(\eta) \ , \ \ A_i^*(\eta) = \frac{\partial_r S_L(\eta)}{\partial(\partial_\eta^r A^i(\eta))}. \tag{5}$$

If we apply the first Noether theorem to the invariance of the density $d\eta S_L(\eta)$ with respect to the $\eta$-shifts by a constant $\mu$ treated as symmetry transformations $(A^i, z^M, \eta) \to (A^i, z^M, \eta+\mu)$, we arrive at the conclusion that the LS has an integral

$$S_E((A, \partial_\eta A)(\eta), \eta) = \frac{\partial_r S_L(\eta)}{\partial(\partial_\eta^r A^i(\eta))} \partial_\eta^r A^i(\eta) - S_L(\eta) : S_H(\eta) = S_E(A(\eta), \partial_\eta A(\Gamma(\eta), \eta), \eta) \tag{6}$$

(which is conserved during the $\eta$-evolution of the quantity) if the equations

$$\frac{\partial_r}{\partial \eta} S_L(\eta) = 0, \ \ (\partial_\eta U_+)(\eta) S_L(\eta)\Big|_{\hat{A}^i(\eta)} = 0 \tag{7}$$

are obeyed for the solutions $\hat{A}^i(\eta)$ of the LS. In terms of the action $S_H(\Gamma(\eta), \eta)$, the *classical Lagrangian master equation* in $\Pi TM_{cl}$ for a solution of the LS [the second equation of (7)] has the form of the *Hamiltonian master equation* for the solution $\hat{\Gamma}(\eta)$ of the HS:

$$\frac{\partial_r}{\partial \eta} S_H(\eta) = 0, \ \ (S_H(\eta), S_H(\eta))_\eta = 0 \ . \tag{8}$$

The functions $\Theta_i(\eta) = \Theta_i((A, \partial_\eta A)(\eta), \eta)$, constraining the statement of the Cauchy problem for the LS, are, in general, functionally dependent. Let us restrict our consideration to an SM representable by natural equations:

$$S_L(\eta) = T(\partial_\eta A(\eta)) - S(A(\eta), \eta) : \ \ \Theta_i(\eta) = -S_{,i}(A(\eta), \eta)(-1)^{\varepsilon_i}, \ \ S_{,i}(\eta) = \frac{\partial_r S(\eta)}{\partial A^i(\eta)}. \tag{9}$$

The equation for the functions $\Theta_i(\eta)$ implies, provided that there exists an $(m_+, m_-)$-dimensional hypersurface $\Sigma \subset M_{cl}$ such that $\Theta_i(\eta)|_\Sigma \equiv 0$, that there are at least $m$ ($m = m_+ + m_-$) independent identities which can be associated with *special gauge transformations* (SGT's) for $A^i(\eta)$:

$$S_{,i}(A(\eta), \eta) R^i_{\alpha_0}(A(\eta), \eta) = 0, \ \ \delta A^i(\eta) = R^i_{\alpha_0}(A(\eta), \eta) \xi_0^{\alpha_0}(\eta) \ , \ \ \alpha_0 = 1, \ldots, m_0 = m_{0+} + m_{0-}, \ \ \varepsilon(\xi_0^{\alpha_0}) = \varepsilon_{\alpha_0}. \tag{10}$$

The generators $R^i_{\alpha_0}(\eta)$ are dependent if the rank $\|R^i_{\alpha_0}(\eta)\|_{|\Sigma} = (m_+, m_-) < (m_{0+}, m_{0-})$, and therefore they have on the surface $\Theta_i(\eta) = 0$ proper null vectors $Z^{\alpha_0}_{\alpha_1}(A(\eta), \eta)$, which, not all independent, exhaust the zero modes of the generators. As a result, the *special type* gauge theory of *L*-stage reducibility is defined by the relations, with $s = 0, \ldots, L-1$ [and with the notation $(m_+, m_-) \equiv (m_{(-1)+}, m_{(-1)-})$],

$$Z^{\alpha_{s-1}}_{\alpha_s}(A(\eta),\eta)Z^{\alpha_s}_{\alpha_{s+1}}(A(\eta),\eta) = S_{,j}(A(\eta),\eta)L^{\alpha_{s-1}j}_{\alpha_{s+1}}(A(\eta),\eta)\ ,\quad \alpha_s = 1,...,m_s = m_{s+} + m_{s-},\quad \varepsilon(Z^{\alpha_s}_{\alpha_{s+1}}) = \varepsilon_{\alpha_s} + \varepsilon_{\alpha_{s+1}},$$

$$(m_{s+},m_{s-}) > \sum_{k=0}^{s}(-1)^k(m_{(s-k-1)+},\ m_{(s-k-1)-}),\quad Z^{\alpha_{-1}}_{\alpha_0}(\eta) \equiv R^i_{\alpha_0}(\eta),\quad L^{\alpha_{-1}j}_{\alpha_1}(\eta) \equiv K^{ij}_{\alpha_1}(\eta) = -(-1)^{\varepsilon_i\varepsilon_j}K^{ji}_{\alpha_1}(\eta)\ ,\quad (11)$$

$$(m_{L+},m_{L-}) = \sum_{k=0}^{L}(-1)^k(m_{(L-k-1)+},\ m_{(L-k-1)-}).$$

The system of projectors on $C^\infty(\Pi TM_{cl})\times\{\eta\}$ : $\{1-\eta\partial_\eta, \eta\frac{\partial}{\partial\eta}, U_+(\eta)\}$ singles out from relations (11) the GA's with structural functions and relations not explicitly depending on $\eta$ and the GA for $\eta = 0$ with standard reducible model relations for $(\varepsilon_P)_I = (\varepsilon_P)_{\alpha_s} = 0$ [2].

Definitions (9)–(11) are also valid for the Hamiltonian formulation of an SM in which the integrability of the HS is guaranteed by that the master equation (8) holds true by virtue of the antibracket Jacobi identity in the expression

$$0 = \partial_\eta^2 \Gamma^p(\eta) = -\big((\Gamma^p(\eta),\ S_H(\Gamma(\eta)))_\eta, S_H(\Gamma(\eta))\big)_\eta = -\tfrac{1}{2}\big(\Gamma^p(\eta),\ (S_H(\Gamma(\eta)),\ S_H(\Gamma(\eta)))_\eta\big)_\eta = 0. \quad (12)$$

This provides validity for the $\eta$-translation formula on $C^\infty(\Pi T^*M_{cl})\times\{\eta\}$

$$\delta_\mu F(\eta)_{|\hat\Gamma} = \mu\partial_\eta F(\eta)_{|\hat\Gamma} = \mu\bar{s}^l(\eta)F(\eta),\quad \bar{s}^l(\eta) = \tfrac{\partial}{\partial\eta} - (S_H(\Gamma(\eta)),\cdot)_\eta \quad (13)$$

and nilpotency for the BRST-like generator of $\eta$-shifts $\bar{s}^l(\eta)$.

**LOCAL SUPERFIELD QUANTIZATION**

The SLQ rules for a standard initial model under a usual ghost number distribution with $gh(A^i) = -gh(A^*_i) - 1 = 0$ [2] and $gh(\eta,\partial_\eta) = (-1,1)$ are initiated by restricting any formulation of the SM to the equations

$$\left(gh,\ \tfrac{\partial}{\partial\eta}\right)S_{H(L)}(\eta) = (0,0). \quad (14)$$

Their nontrivial solution exists if $S_{H(L)}(\eta)$ contains a potential summand $S(A(\eta),0) = S(A(\eta))$: $S(A(0)) = S_0(A)$. The restricted SGT (10), after the substitution $\xi^{\alpha_0}_0(\eta) = C^{\alpha_0}(\eta)d\eta$, can be represented by an almost Hamiltonian system of $2n$ DE's with a Hamiltonian $S^0_1(\Gamma_{cl},C_0)(\eta) = (A^*_i R^i_{\alpha_0}(A)C^{\alpha_0})(\eta)$. The function $S^0_1(\eta)$, by virtue of Eqs. (11), is invariant modulo $S_{,i}(\eta)$ with respect to the new SGT's for the ghosts of $C^{\alpha_0}(\eta)$ with arbitrary functions $\xi^{\alpha_1}_1(\eta)$

$$\delta C^{\alpha_0}(\eta) = Z^{\alpha_0}_{\alpha_1}(A(\eta))\xi^{\alpha_1}_1(\eta)\ ,\quad (\varepsilon,gh)\xi^{\alpha_1}_1(\eta) = \left(\varepsilon_{\alpha_1} + (1,0,1), 1\right), \quad (15)$$

which can be reformulated for $\xi^{\alpha_1}_1(\eta) = C^{\alpha_1}(\eta)d\eta$ as an almost HS with $S^1_1(A,C^*_0,C_1)(\eta) = (C^*_{\alpha_0}Z^{\alpha_0}_{\alpha_1}(A)C^{\alpha_1})(\eta)$.

As a result, all, possibly enhanced, restricted relations (11) for the restricted SM of $L$-stage reducibility are described by a sequence of SGT's for the higher ghosts of $C^{\alpha_{s-1}}(\eta)$ that leave the function $S^{s-1}_1(\eta)$, subject to conditions (14), invariant modulo $S_{,i}(\eta)$ for $s = 1, \ldots, L$:

$$\delta C^{\alpha_{s-1}}(\eta) = Z^{\alpha_{s-1}}_{\alpha_s}(A(\eta))\xi_s^{\alpha_s}(\eta), \quad S_1^{s-1}(\eta) = (C^*_{\alpha_{s-2}}Z^{\alpha_{s-2}}_{\alpha_{s-1}}(A)C^{\alpha_{s-1}})(\eta), \quad (\varepsilon, gh)\xi_s^{\alpha_s} = \left(\varepsilon_{\alpha_s} + s(1,0,1), s\right).^3 \quad (16)$$

The substitution $\xi_s^{\alpha_s}(\eta) = C^{\alpha_s}(\eta)d\eta$ transforms the SGT's (16) into a system of $m_{s-1}$ first-order DE's with respect to $\eta$, extended by introducing $C^*_{\alpha_{s-1}}(\eta)$ to an almost HS with $2m_{s-1}$ DE's in terms of superfields $\Gamma^{p_{s-1}}_{s-1} = (C^{\alpha_{s-1}}, C^*_{\alpha_{s-1}})(\eta)$:

$$\partial^r_\eta \Gamma^{p_{s-1}}_{s-1}(\eta) = \left(\Gamma^{p_{s-1}}_{s-1}(\eta), \ S_1^S(\eta)\right)_\eta. \quad (17)$$

Having combined the systems (17) for all $s$ with the initial HS (4), restricted by condition (14), we obtain an HS nonintegrable in the sense of (12) in $\Pi T^* M_l = \left\{\Gamma^{p_l}_l(\eta) = \left(\Gamma^{p_{cl}}_{cl}, \Gamma^{p_0}_0, ..., \Gamma^{p_L}_L\right)(\eta) = \left(\Phi^{A_l}, \Phi^*_{A_l}\right)(\eta), \quad l = \min\right\}$:

$$\partial^r_\eta \Gamma^{p_l}_l(\eta) = \left(\Gamma^{p_l}_l(\eta), \ S^L_{[1]}(\eta)\right)_\eta, \quad S^L_{[1]}(\eta) = S(A(\eta)) + \sum_{s=0}^{L}\left(C^*_{\alpha_{s-1}}Z^{\alpha_{s-1}}_{\alpha_s}(A)C^{\alpha_s}\right)(\eta) \quad (18)$$

with the function $S^L_{[1]}(\eta)$ satisfying the condition that the $\eta$-local classical master equation has a proper solution in a minimal sector [2].

The integrability of the HS in (18) is provided by the deformation of the function $S^L_{[1]}(\eta)$ in powers of $\Phi^*_{A_l}(\eta)$ and then in powers of $C^{\alpha_s}(\eta)$ in the context of the existence theorem [2] for the solution of the master equation:

$$\left(S_{Hl}(\Gamma_l(\eta)), \ S_{Hl}(\Gamma_l(\eta))\right)_\eta = 0, \quad \left(\varepsilon, \ gh, \ \tfrac{\partial}{\partial\eta}\right)S_{Hl}(\Gamma_l(\eta)) = (\mathbf{0}, 0, 0). \quad (19)$$

The deformation of the function $S_{H\min}(\eta)$ in the Planck constant $\hbar$, which preserves the form of Eq. (19), extended to the proper solution in $\Pi T^* M_l = \left\{\Gamma^{p_l}_l(\eta)\right\}$ (hereinafter $l$ = ext) with pyramids of ghosts and auxiliary superfields [2]

$$\Gamma^{p_l}_l(\eta) = \left(\Gamma^{p_{\min}}_{\min}, C^{\alpha_s}_{s'}, \ B^{\alpha_s}_{s'}, \ C^*_{s'\alpha_s}, \ B^*_{s'\alpha_s}\right)(\eta), \quad s' = 0, ..., s, \quad (\varepsilon, gh)C^{\alpha_s}_{s'} = \left(\varepsilon_{\alpha_s} + (s+1)(1,0,1), 2s' - s - 1\right)$$

$$= (\varepsilon, gh)B^{\alpha_s}_{s'} + ((1,0,1),-1), \quad S_{Hl}(\Gamma_l(\eta)) = S_{H\min}(\Gamma_l(\eta)) + \sum_{s=0}^{L}\sum_{s'=0}^{s}\left(C^*_{s'\alpha_s}B^{\alpha_s}_{s'}\right)(\eta), \quad (20)$$

specifies, for instance, the quantum action $S^\Psi_H(\Gamma(\eta),\hbar) = \exp\left\{(\Psi(\Phi_l(\eta)), \ )_\eta\right\}S_{Hl}(\Gamma_l(\eta),\hbar)$ for the Abelian hypergauge. The action $S^\Psi_H(\eta,\hbar)$ defines an integrable HS, which, in turn, specifies an $\eta$-local not nilpotent generator of BRST transformations in $\Pi T^* M_l$ (standard for $\eta = 0$ in the BV method) which annulls $S^\Psi_H(\eta,\hbar)$ and is associated with its $\eta$-nonintegrable subsystem

$$\tilde{s}^{l(\Psi)}(\eta) = \frac{\partial}{\partial\eta} + \frac{\partial_r S^\Psi_H(\eta,\hbar)}{\partial\Phi^*_{A_l}(\eta)}\frac{\partial_l}{\partial\Phi^{A_l}(\eta)}, \quad \partial^r_\eta\left(\Phi^{A_l}, \Phi^*_{A_l}\right)(\eta) = \left(\left(\Phi^{A_l}(\eta), \ S^\Psi_H(\eta,\hbar)\right)_\eta, \ 0\right). \quad (21)$$

---

[3] In view of Eqs. (14) and $gh(A^i)=0$, we assume hereinafter that $(\varepsilon_P)_i = (\varepsilon_P)_{\alpha_s} = 0$ and, without loss of generality, that the value of $L$ is invariable in this case. In addition, we use the notation $\left(C^{\alpha_{-1}}, C^*_{\alpha_{-1}}\right)(\eta) \equiv \left(A^i, A^*_i\right)(\eta)$.

The Feynman rules, in the general case where the initial superfield–superantifield structure on $\Pi T^* M_{\text{ext}}$ is ignored but the Darboux coordinates $(\varphi, \varphi^*)(\eta)$ are essentially used, are specified by a GFGF $Z(\partial_\eta \varphi^*, \varphi^*, \partial_\eta \varphi, I)(\eta) = Z(\eta)$ having the form of a functional integral with a fixed $\eta$:

$$Z(\eta) = \int d\tilde{\Gamma}_l(\eta) d\lambda(\eta) d\pi(\eta) \exp\left\{\frac{i}{\hbar}\left(S_{Hl}(\tilde{\Gamma}_l(\eta), \hbar) + X^{\Psi_g}\left((\tilde{\varphi}, \tilde{\varphi}^* - \varphi^*, \lambda, \lambda^*, \pi)(\eta)\right)_{|\lambda^* = 0}\right.\right.$$
$$\left.\left. - \left((\partial_\eta \varphi_a^*)\tilde{\varphi}^a + \tilde{\varphi}_a^*(\partial_\eta^r \varphi^a) - I_C \lambda^C - I_D^\pi \pi^D\right)(\eta)\right)\right\}, \qquad (22)$$

with a measure $d\lambda(\eta) = \prod_{t=0}^{K} \prod_{t'=0}^{t} d\lambda_{t'}^t(\eta)$, $d\pi(\eta) = \prod_{t=1}^{K} \prod_{t'=1}^{t} d\pi_{t'}^t(\eta)$. The function $Z(\eta)$ depends on the sources $\left(\partial_\eta \varphi_a^*, \partial_\eta^r \varphi^a, I_C, I_D^\pi\right)(\eta) = \left(-J_a, \lambda^a, I_{0C} + I_{1C}\eta, I_{0D}^\pi + I_{1D}^\pi \eta\right)$, $\left((\vec{\varepsilon}, gh)\partial_\eta \varphi_a^* = (\vec{\varepsilon}, -gh)\varphi^a\right)$ to the superfields $(\varphi^a, \varphi_a^*, \lambda^C, \pi^D)(\eta)$, where $\{\varphi^a(\eta)\} \cap \{\Phi^{A_l}(\eta)\} \neq \{\Phi^{A_l}(\eta)\}$, and $\lambda^C = (\lambda_{t'}^{a_t}, t = 0, \ldots, K, t' = 0, \ldots, t)$, $\pi^D = (\pi_{t'}^{a_t}, t = 1, \ldots, K, t' = 1, \ldots, t)$: $\left(\varepsilon(\pi_{t'}^{a_t}) = \varepsilon(\lambda_{t'}^{a_t}) + (1,0,1) = \varepsilon_{a_t} + (t+1)(1,0,1)\right)$ form pyramids of Lagrangian multipliers to the dependent hypergauges $G_{a_0}(\Gamma_l(\eta))$, $\varepsilon(G_{a_0}) = \varepsilon_{a_0}$, $a_0 = 1, \ldots, k_0 = k_{0+} + k_{0-}$, and to the gauges that remove their degeneracy. The dependence of the hypergauges implies that if the condition $\text{rank}\|\partial G_{a_0}(\eta)/\partial \Gamma_l^{p_l}(\eta)\| \big|_{\partial S/\partial \Gamma = G = 0} = (k_{(-1)+}, k_{(-1)-}) < (k_{0+}, k_{0-})$, $(k_0 > \dim_+ \Pi T^* M_{\text{ext}})$ is fulfilled for the functions $G_{a_0}(\eta)$, there exist proper null vectors $Z_{g\ a_1}^{a_0}(\Gamma_l(\eta))$, $a_1 = 1, \ldots, k_1 = k_{1+} + k_{1-}$, $\varepsilon\left(Z_{g\ a_1}^{a_0}\right) = \varepsilon_{a_0} + \varepsilon_{a_1}$, which exhaust the (not all independent) zero modes of the hypergauges, and so on.

The hypergauge conditions $G_{a_0}(\Gamma_l(\eta))$ are assumed to be solvable with respect to the superantifields $\varphi_a^*(\eta)$. In addition to the involution relations

$$\left(G_{a_0}(\eta), G_{b_0}(\eta)\right)_\eta^{(\Gamma)} = G_{c_0}(\eta) U_{a_0 b_0}^{c_0}(\eta), \quad U_{a_0 b_0}^{c_0}(\eta) = -(-1)^{(\varepsilon_{a_0}+1)(\varepsilon_{b_0}+1)} U_{b_0 a_0}^{c_0}(\eta) \qquad (23)$$

and, perhaps, to the unimodularity relations [9], the functions $G_{a_0}(\eta)$, with the local and $\bar{J}$-covariant descriptions of the theory preserved, specify hypergauge GA's of $K$-stage reducibility by the relations

$$G_{a_0}(\eta) Z_{g\ a_1}^{a_0}(\Gamma_l(\eta)) = 0, \quad Z_{g\ a_t}^{a_{t-1}}(\Gamma_l(\eta)) Z_{g\ a_{t+1}}^{a_t}(\Gamma_l(\eta)) = G_{b_0}(\eta) M_{a_{t+1}}^{a_{t-1}b_0}(\Gamma_l(\eta)), \quad t = 1, \ldots, K-1, \quad a_t = 1, \ldots, k_t = k_{t+} + k_{t-},$$

$$(k_{t+}, k_{t-}) > \sum_{k=0}^{t} (-1)^k (k_{(t-k-1)+}, k_{(t-k-1)-}), \quad (k_{K+}, k_{K-}) = \sum_{k=0}^{K} (-1)^k (k_{(K-k-1)+}, k_{(K-k-1)-}), \qquad (24)$$

where it is taken into account that $\text{rank}\| Z_{g\ a_t}^{a_{t-1}}(\Gamma_l(\eta))\| \big|_{G=0} < (k_{t+}, k_{t-})$. For $K = 0$ the functions $G_{a_0}(\eta)$ are independent [9], i.e., $a_0 \equiv a$.

The quadratic in $(\lambda, \lambda^*, \pi)(\eta)$ part of the gauge fixing action $X((\Gamma_l, \lambda, \lambda^*, \pi)(\eta), \hbar)$,

$$X(\eta,\hbar) = X_{\min}\left((\Gamma_l, \lambda_0, \lambda_0^*)(\eta), \hbar\right) + \sum_{t=1}^{K}\sum_{t'=1}^{t}\left(\lambda_{t'\,a_t}^* \pi_t^{a_t}\right)(\eta),$$
$$X_{\min} = \left(G_{a_0}(\Gamma_l)\lambda_0^{a_0} + \sum_{t=1}^{K}\left(\lambda_{0\,a_{t-1}}^* Z_{g\,a_t}^{a_{t-1}}(\Gamma_l)\lambda_0^{a_t}\right) + O(\lambda^*)\right)$$
(25)

determines the proper solution of one of the systems of equations in $\Pi T^* M_l = \{\Gamma_l(\eta) = (\Gamma_{\text{ext}}, \lambda, \lambda^*, \pi, \pi^*), l=\text{tot}\}$:

1) $(X(\eta,\hbar), X(\eta,\hbar))_\eta^{(\Gamma_l)} = 0$, $\Delta^l(\eta) X(\eta,\hbar) = 0$; 2) $\Delta^l(\eta) \exp\left\{\frac{i}{\hbar} X(\eta,\hbar)\right\} = 0$, (26)

with a nilpotent operator $\Delta^l(\eta)$, $\Delta^l(\eta) = (-1)^{\varepsilon(\Gamma_l^p)} \frac{1}{2} \omega_{p_l q_l}^l(\eta) \left(\Gamma_l^{q_l}(\eta), \left(\Gamma_l^{p_l}(\eta), \phantom{X}\right)_\eta^{(\Gamma_l)}\right)_\eta^{(\Gamma_l)}$. In the functional integral (22), we restrict ourselves to the determination of the Lagrangian surface $\Lambda_g$ parameterized by superfields $(\varphi^*, \lambda, \pi)(\eta)$, such that the action $X^{\Psi_g}(\eta)$ on this surface is nondegenerate and given by the so-called second-level gauge fermion $\Psi_g(\eta)$ [unlike the function $\Psi(\eta)$ in (21)]. The minimal $\Psi_g(\eta)$ may have, for instance, the structure

$$\Psi_g(\Gamma,\lambda)(\eta) = \sum_{t=1}^{K}\sum_{t'=1}^{t}\left(\lambda_{t'}^{a_t}\omega_{a_t b_{t-1}}^{t'}\lambda_{t'-1}^{b_{t-1}}\right)(\eta), \quad \varepsilon(\Psi_g) = (1,0,1),$$ (27)

with constant supermatrices $\left\|\omega_{a_t b_{t-1}}^{t'}\right\|$, which provide nondegeneracy of the restriction on $\Lambda_g$ for the function

$$X^{\Psi_g}(\eta) = \exp\left\{\left(\Psi_g(\eta), \phantom{X}\right)_\eta\right\} X(\eta,\hbar).$$ (28)

For $\partial_\eta \varphi^* = I = 0$, the integrand in Eq. (22) is invariant under the *superfield BRST transformations* $\tilde{\Gamma}_{\text{tot}}(\eta) = (\tilde{\Gamma}_{\text{ext}}, \lambda, \lambda^*, \pi, \pi^*)(\eta) \to \tilde{\Gamma}_{\text{tot}}(\eta) + \delta_\mu \tilde{\Gamma}_{\text{tot}}(\eta)$, which represent $\eta$-shifts by the odd parameter $\mu$ along the solution $\breve{\Gamma}_{\text{tot}}(\eta)$ of the $\eta$-nonintegrable almost HS

$$\begin{cases} \partial_\eta^r \tilde{\Gamma}_{\text{ext}}^{P_{\text{ext}}}(\eta) = \left(\tilde{\Gamma}_{\text{ext}}^{P_{\text{ext}}}(\eta), \left(S - X^{\Psi_g}\right)(\eta)\right)_\eta^{(\tilde{\Gamma})} \Big|_{\lambda^*=0}, \\ \partial_\eta^r \lambda^C(\eta) = -2\left(\lambda^C(\eta), X^{\Psi_g}(\eta)\right)_\eta \Big|_{\lambda^*=0}, \\ \partial_\eta^r \pi^D(\eta) = -2\left(\pi^D(\eta), X^{\Psi_g}(\eta)\right)_\eta \Big|_{\lambda^*=0}, \\ \partial_\eta^r (\lambda^*, \pi^*, \varphi^*)(\eta) = 0. \end{cases} \Leftrightarrow \delta_\mu \tilde{\Gamma}_{\text{tot}}^{P_{\text{tot}}}(\eta) = \partial_\eta^r \tilde{\Gamma}_{\text{tot}}^{P_{\text{tot}}}(\eta)\mu \Big|_{\breve{\Gamma}_{\text{tot}}}.$$ (29)

In this case, the actions $S(\eta) = S_{H\text{ext}}(\tilde{\Gamma}_{\text{ext}}(\eta), \hbar)$ and $X^{\Psi_g}\left((\tilde{\varphi}, \tilde{\varphi}^* - \varphi^*, \lambda, \lambda^*, \pi)(\eta)\right)$ may satisfy different but fixed systems in (26).

The HS (29) permits one to establish the independence of the vacuum functional $Z_X(\varphi^*) = Z(\varphi^*, 0, 0, 0)$ and, hence, the $S$-matrix, in virtue of the equivalence theorem [11], on the choice of the gauge. Actually, the $\eta$-shift along $\breve{\Gamma}_{\text{tot}}(\eta)$ by $\mu\left((\tilde{\varphi}, \tilde{\varphi}^* - \varphi^*, \lambda, \lambda^*, \pi)(\eta)\right)$ corresponding to the HS (29) in $Z_{X+\Delta X}(\varphi^*)$ is compensated by an additional $\eta$-shift by a constant $\mu_1$ along the solution of an almost HS of the form (29) with a Hamiltonian $W\left((\tilde{\varphi}, \tilde{\varphi}^* - \varphi^*, \lambda, \lambda^*, \pi)(\eta), \hbar\right)$: $\left(\varepsilon, \frac{\partial}{\partial \eta}\right) W(\eta) = (\mathbf{0}, 0)$ related to $\mu(\eta)$ by $W(\eta)\mu_1 = -i\hbar\mu(\eta)$. As a result, we have

$$Z_{X+\Delta X}(\varphi^*(\eta)) = \int d\tilde{\Gamma}_{ext}(\eta) d\lambda(\eta) d\pi(\eta) \exp\left\{\frac{i}{\hbar}\left(S + \left(X^{\Psi_g} + \Delta X^{\Psi_g} + \Delta \tilde{X}\right)_{|\lambda^* = \pi^* = 0}\right)(\eta)\right\}. \tag{30}$$

If the action $\left(X^{\Psi_g} + \Delta X^{\Psi_g}\right)(\eta)$ obeys the system in (26) that is also valid for the function $X^{\Psi_g}(\eta)$, the variation $\Delta X^{\Psi_g}(\eta)$ satisfies a linear equation with a nilpotent operator $Q_j(X)$:

$$Q_j(X)\Delta X(\eta) = 0, \quad Q_j(X) = \left(X^{\Psi_g}(\eta), \quad \right)_\eta - \delta_{j2}(i\hbar\Delta(\eta)), \quad j = 1, 2. \tag{31}$$

As in [12], the general solution of this equation, vanishing for $\Gamma_{tot}(\eta) = 0$ has the form

$$\Delta X^{\Psi_g}(\eta) = Q_j(X)\Delta Y(\eta), \quad \left(\varepsilon, \tfrac{\partial}{\partial \eta}\right)\Delta Y(\eta) = ((1,0,1), 0). \tag{32}$$

Assuming that the action $W(\eta)$ satisfies the same system as that for the function $X^{\Psi_g}(\eta)$ and taking account of the representation for $\Delta \tilde{X}(\eta)$, $\Delta \tilde{X}(\eta) = -2Q_j(X)W(\eta)\mu_1$, we obtain coincidence of $Z_{X+\Delta X}(\varphi^*)$ with $Z_X(\varphi^*)$ by setting $W(\eta)\mu_1 = \tfrac{1}{2}\Delta Y(\eta)$.

Another implication of the corresponding systems in Eqs. (26) for $S(\eta)$ and $X^{\Psi_g}(\eta)$ is the Ward identity for the GFGF $Z(\eta)$

$$\left(\left\{\partial_\eta \varphi_a^*(\eta) \frac{\partial_l}{\partial \varphi_a^*(\eta)} - \left(\frac{\partial S_{Hext}}{\partial \tilde{\varphi}^a(\eta)}\right)\left(i\hbar \frac{\partial_l}{\partial(\partial_\eta \varphi^*)}, i\hbar \frac{\partial_r}{\partial(\partial_\eta^r \varphi)}\right)(\eta)\right\} \frac{\partial_l}{\partial \varphi_a^*(\eta)} \right. \tag{33}$$
$$\left. + \frac{i}{\hbar} I_C(\eta) \frac{\partial_l X^{\Psi_g}\left(\left(i\hbar \frac{\partial_l}{\partial(\partial_\eta \varphi^*)}, i\hbar \frac{\partial_r}{\partial(\partial_\eta^r \varphi)}, -\varphi^*, -i\hbar \frac{\partial_l}{\partial I}, \lambda^*, -i\hbar \frac{\partial_l}{\partial I^\pi}\right)(\eta)\right)}{\partial \lambda_C^*(\eta)}\bigg|_{\lambda^* = 0} z(\eta) = 0\right)$$

which is obtained by functional averaging, for instance, of the second equation in (26) for $X^{\Psi_g}\left((\tilde{\varphi}, \tilde{\varphi}^* - \varphi^*, \lambda, \lambda^*, \pi)(\eta)\right)$ (provided that $X^{\Psi_g}(\eta)$ is the solution of the same equation)

$$\int d\tilde{\Gamma}_{ext}(\eta) d\lambda(\eta) d\pi(\eta) \exp\left\{\tfrac{i}{\hbar}\left(S_{Hext}\left(\tilde{\Gamma}_{ext}(\eta), \hbar\right) - \left(\left(\partial_\eta \varphi_a^*\right)\tilde{\varphi}^a + \tilde{\varphi}_a^*\left(\partial_\eta^r \varphi^a\right) - I_C \lambda^C - I_D^\pi \pi^D\right)(\eta)\right)\right\} \times$$
$$\Delta(\eta) \exp\left\{\tfrac{i}{\hbar} X^{\Psi_g}\left((\tilde{\varphi}, \tilde{\varphi}^* - \varphi^*, \lambda, \lambda^*, \pi)(\eta)\right)\right\}\bigg|_{\lambda^* = 0} = 0, \tag{34}$$

integration by parts in (34), and use of the rule $\left(\frac{\partial}{\partial \varphi^*} - \frac{\partial}{\partial \tilde{\varphi}^*}\right) X^{\Psi_g}(\eta) = 0$.

The definition of GFGF $Z(\eta)$ in the form of (22) suffices to introduce the effective action $\Gamma(\eta) = \Gamma(\varphi, \varphi^*, \partial_\eta \varphi, I)(\eta)$ by means of the Legendre transformation of $Z(\eta)$ with respect to $\partial_\eta \varphi^*(\eta)$:

$$\Gamma(\eta) = \frac{\hbar}{i}\ln \mathcal{Z}(\eta) + \left(\left(\partial_\eta \varphi_a^*\right)\varphi^a\right)(\eta), \qquad \varphi^a(\eta) = -\frac{\hbar}{i}\frac{\partial_l \ln Z(\eta)}{\partial(\partial_\eta \varphi_a^*(\eta))}. \tag{35}$$

The Ward identity for $\Gamma(\eta)$ with $\Gamma''_{ab}(\eta) = \frac{\partial_l}{\partial \varphi^a(\eta)}\frac{\partial_r}{\partial \varphi^b(\eta)}\Gamma(\eta)$ $\left(\Gamma''_{ba}(\eta)\left(\Gamma''^{-1}\right)^{ac} = \delta_b{}^c\right)$ has the form

$$\begin{aligned}
I_C(\eta)&\frac{\partial_l X^{\Psi_g}\left(\varphi^a + i\hbar\left(\Gamma''^{-1}\right)^{ac}\frac{\partial_l}{\partial \varphi^c(\eta)}, i\hbar\frac{\partial_r}{\partial(\partial_\eta^r \varphi)} - \frac{\partial_r \Gamma}{\partial(\partial_\eta^r \varphi)} - \varphi^*, \frac{\partial_l \Gamma}{\partial I} - i\hbar\frac{\partial_l}{\partial I}, \lambda^*, \frac{\partial_l \Gamma}{\partial I^\pi} - i\hbar\frac{\partial_l}{\partial I^\pi}\right)(\eta)}{\partial \lambda_C^*(\eta)}\bigg|_{\lambda^*=0} \\
&-\left(\frac{\partial S_{H\text{ext}}}{\partial \tilde{\varphi}^a(\eta)}\right)\left(\left(\varphi^b + i\hbar\left(\Gamma''^{-1}\right)^{bc}\frac{\partial_l}{\partial \varphi^c(\eta)}, i\hbar\frac{\partial_r}{\partial(\partial_\eta^r \varphi)} - \frac{\partial_r \Gamma}{\partial(\partial_\eta^r \varphi)}\right)(\eta)\right)\frac{\partial_l \Gamma(\eta)}{\partial \varphi_a^*(\eta)} + \frac{1}{2}(\Gamma(\eta), \Gamma(\eta))_\eta = 0.
\end{aligned} \tag{36}$$

For $I_C(\eta) = 0$, identities (33) and (36) take the standard form, complicated by the presence of the sources $(\partial_\eta^r \varphi^a, I_D^\pi)(\eta)$, but for the generating functionals $\mathcal{Z}(\eta)$ and $\Gamma(\eta)$ in non-Abelian hypergauges.

**ASSOCIATION WITH THE BV METHOD AND SUPERFIELD LAGRANGIAN QUANTIZATION**

For a special type SM, the association of the quantities of a local SLQ with the standard description of field theory objects in the framework of the BV method [2] is provided by the restriction $\eta = 0$ for all the quantities and relations, for instance: $\left(M, \Pi T^* M_{\text{tot}}, I_C, I_D^\pi\right) \to \left(\tilde{M}, \Pi T^* M_{tot}\big|_{\eta=0} = \left\{\Gamma_0^{p_{tot}}{}_{tot}\right\}, I_{0C}, I_{0D}^\pi\right)$. Additionally, for the SM given by expressions (3)–(8), one must eliminate from $S_L(\eta)$ and $S_H(\eta)$ the superfields $\partial_\eta A^i(\eta)$ and $A_i^*(\eta)$ and those superfields of $A^i(\eta)$ which contain functions with a wrong relationship between the spin and the statistic, $gh(A^i) \neq 0$ and $(\varepsilon_P)_i \neq 0$, either by setting them equal to zero or by applying horizontality conditions, as in the case of Yang–Mills theories [13, 14].

For an arbitrary smooth functional $F[\Gamma_l]$, which is used in the methods described in [4, 5], there exists a function $\mathcal{F}(\eta) = \mathcal{F}((\Gamma_l, \partial_\eta \Gamma_l)(\eta), \eta)$, $l = \{cl, \min, ext\}$ such that

$$F[\Gamma_l] = \int d\eta \ \eta \ \mathcal{F}(\eta) = \mathcal{F}(\Gamma_l(0), \partial_\eta \Gamma_l, 0) = \mathcal{F}(\Gamma_{0l}, \Gamma_{1l}), \tag{37}$$

and $F[\Gamma_l]$ does not depend on $\Gamma_{1l}^p$ if $\mathcal{F}(\eta) = \mathcal{F}((\Gamma_l(\eta), \eta)$. From Eq. (37) it follows that there exists a relationship between the local operations $(\bullet, \bullet)_\eta$, $\Delta(\eta)$ and the functional operations $(\bullet, \bullet)$, $\Delta$ [4], which coincide with their analogs in the BV method:

$$\left((F(\eta), G(\eta))_\eta, \ \Delta(\eta)F(\eta)\right)\bigg|_{\eta=0} = \left((F[\Gamma_l], G[\Gamma_l]), \ \Delta F[\Gamma_l]\right). \tag{38}$$

For $l = \text{ext}$, the action of the operators $U$ and $V$ [4] coincides with the action of $(\partial_\eta U_+)(0)$, $(\partial_\eta V_+)(0) = \partial_\eta \Phi^*_{A_l}(\eta)\frac{\partial_l}{\partial \Phi^*_{A_l}(0)}$.

Restricting ourselves to the case of independent hypergauges, we obtain, in view of the representation of the action $X(\eta)$ in (25), that for $\varphi_0^* = 0$ and $\eta = 0$ the vacuum functional $\mathcal{Z}_X(\varphi_0^*)$ coincides with the "first level" functional integral $Z_1$ [9] with the density $M(\Gamma_0) = 1$ determining the measure $\mu[\Gamma_0] = M d\Gamma_0$,

$$Z_1 = \int d\Gamma_0 d\lambda_0 \exp\left\{\tfrac{i}{\hbar}\left(S(\Gamma_0) + G_a(\Gamma_0)\lambda_0^a\right)\right\}. \tag{39}$$

If the functions $S(\eta)$ and $X(\eta)$ satisfy the second system in (26), the transformations (29), for $\pi^D = \pi^*_D = \eta = 0$, coincide, up to the sign, with the BRST transformations for $Z_1$ [9]. Simultaneously, the quantities $S[\Gamma] = S_{Hl}(\Gamma_{0l}) - (\partial_\eta \Phi^*_{A_l})\phi^{A_l}$, and $\bar{X}[\Gamma] = X(\Gamma_{0l}, \lambda_0) - \phi^*_{A_l} \partial^r_\eta \Phi^{A_l}(\eta)$, $l = ext$, satisfy the generating equations given in [4, 5] having the form of the second system in (26) with the operators $\Delta_1 = \Delta - (i\hbar)^{-1}V$, $\Delta_2 = \Delta + (i\hbar)^{-1}U$ defined, respectively, in $\Pi T^* M_{ext}$ and on the surface $\lambda^{A_l}_0(0) = \partial^r_\eta \Phi^{A_l}(\eta)$ for $K = 0$. In the particular case of the boundary condition for $\bar{X}[\Gamma]$ with $\varphi^a_0 = \phi^{A_l}$

$$\frac{\delta}{\delta \lambda^{A_l}}\left(\bar{X}[\Gamma] + \phi^*_{B_l}\lambda^{B_l}\right) = G_{A_l}(\Gamma_l) \tag{40}$$

and with $S[\Gamma] = S(\phi, \phi^*, J)$ independent on the fields $\lambda^{A_l}$, this provides coincidence of $Z_X(\varphi^*_0)$ with $\varphi^*_0 = 0$ with the vacuum functional Z of [5] written in component form:

$$Z = \int d\phi d\phi^* d\lambda \exp\left\{\frac{i}{\hbar}\left(S(\phi,\phi^*,J) + \bar{X}(\phi,\phi^*,\lambda,J) + \partial_\eta\left(\Phi^*_{A_l}\Phi^{A_l}\right)(\eta)\right)_{|J=0}\right\}. \tag{41}$$

This correspondence permits one to state that the effective action can also be determined in the framework of the method described in [4, 5].

## ACKNOWLEDGEMENT

The author is grateful to P. M. Lavrov for useful critical comments.